\documentclass[prl,twocolumn,groupedaddress,showpacs,nofootinbib]{revtex4}
\usepackage{graphicx}
\usepackage{dcolumn}
\usepackage{amssymb}
\usepackage{mathrsfs}
\usepackage{amsmath}
\usepackage{epsfig}
\usepackage[dvips]{color}
\usepackage{hhline}

\begin{document}

\title{Mirror left-right symmetry}

\author{Pei-Hong Gu}
\email{peihong.gu@mpi-hd.mpg.de}

\affiliation{Max-Planck-Institut f\"{u}r Kernphysik, Saupfercheckweg
1, 69117 Heidelberg, Germany}

\begin{abstract}

We propose a novel $SU(3)_c^{}\times SU(2)_L^{}\times SU(2)_R^{}
\times U(1)_{B-L}^{}$ left-right symmetric model where the standard
model fermion and Higgs fields are $SU(2)_L^{}$ doublets or $SU(2)$
singlets while their mirror partners are $SU(2)_R^{}$ doublets or
$SU(2)$ singlets. The scalar fields also include a real singlet for
dark matter and two $SU(2)$ triplets for seesaw. The mixing between
the standard model and mirror fermions is forbidden by a
$Z_2^{}\times Z'^{}_2$ discrete symmetry. The mirror charged
fermions can decay into their standard model partners with the
dark-matter scalar while the mirror neutrinos can decay into the
mirror charged fermions through the right-handed gauge interactions.
Our model can have new implications on the strong CP problem,
leptogenesis, collider phenomenology and dark matter detection.

\end{abstract}


\maketitle

The $SU(3)_c^{}\times SU(2)_L^{}\times SU(2)_R^{}\times
U(1)_{B-L}^{}$ left-right symmetric theory \cite{ps1974}, motivated
by restoring the parity invariance, has become one of the most
attractive proposals beyond the $SU(3)_c^{}\times SU(2)_L^{}\times
U(1)_{Y}^{}$ standard model (SM). In the existent left-right
symmetric models, the SM left-handed fermions are placed in the
$SU(2)_L^{}$ doublets as they are in the SM while the SM
right-handed fermions plus the right-handed neutrinos are placed in
the $SU(2)_R^{}$ doublets. After the left-right symmetry is
spontaneously broken down to the electroweak symmetry, we can obtain
the lepton number violation for the seesaw \cite{minkowski1977}
mechanism. In the seesaw context \cite{minkowski1977,mw1980}, the
leptogenesis \cite{fy1986} for generating the baryon asymmetry in
the universe can be realized by the Yukawa and scalar interactions
\cite{fy1986,mz1992}. The left-right symmetric model for the
universal seesaw scenario \cite{berezhiani1983}, where additional
$SU(2)$-singlet fermions with heavy masses are introduced to
construct the Yukawa couplings with the $SU(2)$-doublet fermions and
Higgs, can provide a solution to the strong CP problem if the parity
symmetry is imposed \cite{bm1989}.

In this paper we shall realize the left-right symmetry in a new way
where the SM fermion and Higgs fields are the $SU(2)_L^{}$ doublets
or $SU(2)$ singlets while their mirror \cite{ly1956} partners are
the $SU(2)_R^{}$ doublets or $SU(2)$ singlets. As a consequence of
the parity symmetry, we can know the ratio among the mirror charged
fermion masses from the SM charged fermion mass spectrum.
Furthermore, the mirror right-handed neutrinos and the SM
left-handed neutrinos can have the Majorana mass matrices with a
same texture through the type-II seesaw \cite{mw1980}. The parity
can also guarantee a vanishing strong CP phase from the
$\theta$-vacuum of the QCD and the mass matrices of the SM and
mirror quarks. The mirror charged fermions can directly decay into
their SM partners with a dark-matter scalar while the mirror
neutrinos can decay into the lighter mirror charged fermions by
exchanging the right-handed charged gauge boson. So, we can have a
leptogenesis with the CP asymmetry formulated by the SM neutrino
masses, lepton and quark mixing matrices. For a left-right symmetry
breaking at $\mathcal{O}(10^{8}_{}\,\textrm{GeV})$, the
first-generation mirror charged fermions can be verified by the
ongoing and future colliders. In this case the dark-matter scalar
can be found at the colliders even if it is not sensitive to the
dark matter direct detection experiments.

\begin{table}
\begin{center}
\begin{tabular}{|c|c|c|}  \hline &&\\[-2.5mm]
$\textrm{Field}$ &$\begin{array}{l}SU(3)_{c}^{}\times
SU(2)_{L}^{}\times \\
SU(2)_{R}^{}\times U(1)_{B-L}^{}\end{array}$
 & $Z^{}_{2}\times
Z'^{}_{2}$~\\
&&\\[-3.0mm]
\hline \hline&&\\[-2.5mm]$\begin{array}{c}q_{L}^{}=
\left[\begin{array}{c}u_{L}^{}\\
[0mm]d_{L}^{}\end{array}\right]\\[-1mm]
-\,-\,-\,-\,-\,-\,-\,-\\[-1mm]
q'^{}_{R}=
\left[\begin{array}{c}u'^{}_{R}\\
[0mm]d'^{}_{R}\end{array}\right]\end{array}$ &
$\begin{array}{c}(\textbf{3},~~\textbf{2},~~\textbf{1},\,+\frac{1}{3})\\
-\,-\,-\,-\,-\,-\,-\,-\\
(\textbf{3},~~\textbf{1},~~\textbf{2},\,+\frac{1}{3})\end{array}$
& $\begin{array}{c}(-,+)\\
-\,-\,-\\
(+,-)\end{array}$\\
&& \\[-3mm]
\hline &&\\[-2.5mm]$\begin{array}{c}d_{R}^{}\\[-1mm]
-\,-\,-\,-\,-\,-\,-\,-\\[-1mm]
d'^{}_{L}\end{array}$ &
$\begin{array}{c}(\textbf{3},~~\textbf{1},~~\textbf{1},\,-\frac{2}{3})\\
-\,-\,-\,-\,-\,-\,-\,-\\(\textbf{3},~~\textbf{1},~~\textbf{1},\,-\frac{2}{3})\end{array}$
& $\begin{array}{c}(-,+)\\
-\,-\,-\\
(+,-)\end{array}$\\
&& \\[-2.7mm]
\hline &&\\[-2.5mm]$\begin{array}{c}u_{R}^{}\\[-1mm]
-\,-\,-\,-\,-\,-\,-\,-\\[-1mm]
u'^{}_{L}\end{array}$ &
$\begin{array}{c}(\textbf{3},~~\textbf{1},~~\textbf{1},\,+\frac{4}{3})\\
-\,-\,-\,-\,-\,-\,-\,-\\(\textbf{3},~~\textbf{1},~~\textbf{1},\,+\frac{4}{3})\end{array}$
& $\begin{array}{c}(-,+)\\
-\,-\,-\\
(+,-)\end{array}$\\
&& \\[-2.7mm]
\hline && \\[-2.7mm]$\begin{array}{c}l_{L}^{}=
\left[\begin{array}{c}\nu_{L}^{}\\
[0mm]e_{L}^{}\end{array}\right]\\[-0.7mm]
-\,-\,-\,-\,-\,-\,-\,-\\[-1mm]
l'^{}_{R}=
\left[\begin{array}{c}\nu'^{}_{R}\\
[0mm]e'^{}_{R}\end{array}\right]\end{array}$ &
$\begin{array}{c}(\textbf{1},~~\textbf{2},~~\textbf{1},\,-1)\\[1mm]
-\,-\,-\,-\,-\,-\,-\,-\\[1mm]
(\textbf{1},~~\textbf{1},~~\textbf{2},\,-1)\end{array}$ &
$\begin{array}{c}(-,+)\\[1mm]
-\,-\,-\\[1mm]
(+,-)\end{array}$\\
&& \\[-3mm]
\hline&& \\[-2.5mm]$\begin{array}{c}e_{R}^{}\\[-1mm]
-\,-\,-\,-\,-\,-\,-\,-\\[-1mm]
e'^{}_{L}\end{array}$ &
$\begin{array}{c}(\textbf{1},~~\textbf{1},~~\textbf{1},\,-2)\\[-0.5mm]
-\,-\,-\,-\,-\,-\,-\,-\\[-0.5mm](\textbf{1},~~\textbf{1},~~\textbf{1},\,-2)\end{array}$
& $\begin{array}{c}(-,+)\\[-0.5mm]
-\,-\,-\\[-0.5mm]
(+,-)\end{array}$\\
&& \\[-2.7mm]
\hline \hline&&\\[-3mm]$ \begin{array}{c}\phi_L^{}=
\left[\begin{array}{c}\phi_{L}^{0}\\
[1mm]\phi_{L}^{-}\end{array}\right]\\[-1mm]
-\,-\,-\,-\,-\,-\,-\,-\\[-1mm]
\phi_R^{}=
\left[\begin{array}{c}\phi_{R}^{0}\\
[1mm]\phi^{-}_{R}\end{array}\right]\end{array}$ &
$\begin{array}{c}(\textbf{1},~~\textbf{2},~~\textbf{1},\,-1)\\[1.5mm]
-\,-\,-\,-\,-\,-\,-\,-\\[1.5mm](\textbf{1},~~\textbf{1},~~\textbf{2},\,-1)\end{array}$
& $\begin{array}{c}(+,+)\\[1.5mm]
-\,-\,-\\[1.5mm]
(+,+)\end{array}$\\
&& \\[-3mm]
\hline &&\\[-2.5mm]$ \begin{array}{c}\xi_L^{}=
\left[\begin{array}{cc}\frac{1}{\sqrt{2}}\xi_{L}^+&\xi_L^{++}\\
[1mm]\xi_{L}^0&-\frac{1}{\sqrt{2}}\xi_{L}^+\end{array}\right]\\[-0.5mm]
-\,-\,-\,-\,-\,-\,-\,-\,-\,-\\[-0.5mm]
\xi_R^{}=
\left[\begin{array}{cc}\frac{1}{\sqrt{2}}\xi_{R}^+&\xi_R^{++}\\
[1mm]\xi_{R}^0&-\frac{1}{\sqrt{2}}\xi_{R}^+\end{array}\right]\end{array}$
&
$\begin{array}{c}(\textbf{1},~~\textbf{3},~~\textbf{1},\,+2)\\[2mm]
-\,-\,-\,-\,-\,-\,-\,-\\[2mm](\textbf{1},~~\textbf{1},~~\textbf{3},\,+2)\end{array}$
&  $\begin{array}{c}(+,+)\\[2mm]
-\,-\,-\\[2mm]
(+,+)\end{array}$\\
&& \\[-2.7mm]
\hline&&\\[-2.5mm]$\chi$ &
$(\textbf{1},~~\textbf{1},~~\textbf{1},\,~~0)$  & $(-,-)$\\
&& \\[-2.7mm]
\hline \hline &&\\[-2.5mm]$G^a_\mu$ &
$(\textbf{8},~~\textbf{1},~~\textbf{1},\,~~0)$  & $(+,+)$\\
&& \\[-2.7mm]
\hline &&\\[-2.5mm]$\begin{array}{c}W^a_{L_\mu^{}}\\[-0.5mm]
-\,-\,-\,-\,-\,-\,-\,-\\[-0.5mm]
W^a_{R_\mu^{}}\end{array}$ &
$\begin{array}{c}(\textbf{1},~~\textbf{3},~~\textbf{1},\,~~0)\\[-0.5mm]
-\,-\,-\,-\,-\,-\,-\,-\\[-0.5mm](\textbf{1},~~\textbf{1},~~\textbf{3},\,~~0)\end{array}$
& $\begin{array}{c}(+,+)\\[-0.5mm]
-\,-\,-\\[-0.5mm]
(+,+)\end{array}$\\
&& \\[-2.7mm]
\hline &&\\[-2.5mm]$B^{}_\mu$ &
$(\textbf{1},~~\textbf{1},~~\textbf{1},\,~~0)$  & $(+,+)$
\\[1mm]\hline
\end{tabular}
\caption{\label{field} Field content. }
\end{center}
\end{table}

The field content is summarized in Table \ref{field}. Under the
parity symmetry, the fields transform as
\begin{eqnarray}
&q_{L}^{}\leftrightarrow q'^{}_{R}\,, ~d_{R}^{}\leftrightarrow
d'^{}_{L}\,,~u_{R}^{}\leftrightarrow
u'^{}_{L}\,,~l_{L}^{}\leftrightarrow l'^{}_{R}\,,
~e_{R}^{}\leftrightarrow
e'^{}_{L}\,,&\nonumber\\
&\xi_L^{}\leftrightarrow\xi^{}_R\,,~\phi_L^{}\leftrightarrow
\phi^{}_R\,,~\chi\leftrightarrow\chi\,,&\nonumber\\
&G^a_\mu\leftrightarrow G^a_\mu\,,~W^a_{L_\mu^{}}\leftrightarrow
W^a_{R_\mu^{}}\,,~B_\mu^{}\leftrightarrow B_\mu^{}\,.&
\end{eqnarray}
The kinetic terms are
\begin{eqnarray}
\mathcal{L}_K^{}&=&\!\!\begin{array}{l}-\frac{1}{4}G_{\mu\nu}^{a}G^{a\mu\nu}_{}-\frac{1}{4}W_{L(R)_{\mu\nu}^{}}^{a}W^{a\mu\nu}_{L(R)}
-\frac{1}{4}B_{\mu\nu}^{}B^{\mu\nu}_{}\end{array}\nonumber\\
&&+\textrm{Tr}\{[D_\mu^{}\xi_{L(R)}^{}]^\dagger_{}D^\mu_{}\xi_{L(R)}^{}\}
+[D_\mu^{}\phi_{L(R)}^{}]^\dagger_{}D^\mu_{}\phi_{L(R)}^{}\nonumber\\
&&\!\!\begin{array}{l}
+\frac{1}{2}\partial_\mu^{}\chi\partial^\mu_{}\chi \end{array}\!\!+
i\bar{q}_{L}^{}\!\not\!\!D q_{L}^{}+i\bar{q}'^{}_{R}\!\not\!\!D
q'^{}_{R}+i\bar{d}_{R}^{}
\!\not\!\!D d_{R}^{}\nonumber\\
&&+i\bar{d}'^{}_{L} \!\not\!\!D d'^{}_{L}+i\bar{u}_{R}^{}
\!\not\!\!D u_{R}^{}+i\bar{u}'^{}_{L}\! \not\!\!D
u'^{}_{L}+i\bar{l}_{L}^{} \!\not\!\!D
l_{L}^{}\nonumber\\
&&+ i \bar{l}'^{}_{R} \!\not\!\!D l'^{}_{R}+i\bar{e}_{R}^{}
\!\not\!\!D e_{R}^{} +i\bar{e}'^{}_{L} \!\not\!\!D e'^{}_{L}\,,
\end{eqnarray}
where the $SU(2)$ doublets and singlets have the covariant
derivatives,
\begin{eqnarray}
\begin{array}{l}D_\mu^{}=\partial_\mu^{}+i g_1^{}\frac{B-L}{2}
B_\mu^{}+ig_2^{}\frac{\tau_a^{}}{2}W_{L_\mu^{}(R_\mu^{})}^a+i
g_3^{}\frac{\lambda_a^{}}{2}G_\mu^a\,,\end{array}
\end{eqnarray}
while the $SU(2)$-triplet Higgs scalars have
\begin{eqnarray}
D_\mu^{}\xi_{L(R)}^{}=(\partial_\mu^{}\!+\!i g_1^{}
B_\mu^{})\xi_{L(R)}^{}\!+\!\!\begin{array}{l}i g_2^{}
W_{L_\mu^{}(R_\mu^{})}^a[\frac{\tau_a^{}}{2},\xi_{L(R)}^{}]\end{array}\!.\nonumber\quad\quad
\end{eqnarray}
\vspace{-1.35cm}
\begin{eqnarray}
\end{eqnarray}
The full scalar potential is
\begin{eqnarray}
\label{potential}
V&=&\mu_{\xi_{L(R)}^{}}^2\textrm{Tr}[\xi_{L(R)}^\dagger\xi_{L(R)}^{}]
+\lambda_\xi^{}\{\textrm{Tr}[\xi_{L(R)}^\dagger
\xi_{L(R)}^{}]\}^2_{}\nonumber\\
&&+\lambda'^{}_\xi\textrm{Tr}[\xi_{L(R)}^\dagger
\xi_{L(R)}^\dagger]\textrm{Tr}[\xi_{L(R)}^{}
\xi_{L(R)}^{}]\nonumber\\
&&+\lambda''^{}_\xi \textrm{Tr}(\xi_L^\dagger
\xi_L^{})\textrm{Tr}(\xi_R^\dagger \xi_R^{})+\mu_{\phi_{L(R)}^{}}^2
\phi_{L(R)}^\dagger \phi_{L(R)}^{}\nonumber\\
&&+\lambda_\phi^{}[\phi_{L(R)}^\dagger
\phi_{L(R)}^{}]^2_{}+2\lambda'^{}_\phi\phi_L^\dagger\phi_L^{}\phi_R^\dagger\phi_R^{}
+\!\!\begin{array}{l}\frac{1}{2}\mu_{\chi}^2
\chi^2_{}\end{array}\nonumber\\
&&+\!\!\begin{array}{l}\frac{1}{4}\lambda_{\chi}^{}\chi^4_{}\end{array}+2\lambda_{\phi\xi}^{}\phi_{L(R)}^\dagger
\phi_{L(R)}^{}\textrm{Tr}[\xi_{L(R)}^\dagger
\xi_{L(R)}^{}]\nonumber\\
&&+2\lambda'^{}_{\phi\xi}\phi_{L(R)}^\dagger
\phi_{L(R)}^{}\textrm{Tr}[\xi_{R(L)}^\dagger
\xi_{R(L)}^{}]\nonumber\\
&&+\rho_{\xi_{L(R)}^{}}^{}[\phi_{L(R)}^T i\tau_2^{} \xi_{L(R)}^{}
\phi_{L(R)}^{} +
\textrm{H.c.}]\nonumber\\
&&+\lambda_{\chi\xi}^{}\chi^2_{}\textrm{Tr}[\xi_{L(R)}^\dagger
\xi_{L(R)}^{}]+\lambda_{\chi\phi}^{}\chi^2_{}\phi_{L(R)}^\dagger
\phi_{L(R)}^{}
\end{eqnarray}
without any complex parameters. The allowed Yukawa interactions only
include
\begin{eqnarray}
\label{yukawa}
\mathcal{L}_Y^{}=-y_{d}^{}(\bar{q}_{L}^{}\tilde{\phi}_{L}^{}
d_{R}^{} + \bar{q}'^{}_{R}\tilde{\phi}_R^{}
d'^{}_{L})-y_{u}^{}(\bar{q}_{L}^{}\phi_L^{} u_{R}^{} +
\bar{q}'^{}_{R}\phi_R^{}
u'^{}_{L})&&\nonumber\\
-y_{e}^{}(\bar{l}_{L}^{}\tilde{\phi}_L^{} e_{R}^{} +
\bar{l}'^{}_{R}\tilde{\phi}_R^{} e'^{}_{L})-
\chi(h_d^{}\bar{d}_R^{}d'^{}_L+h_u^{}\bar{u}_R^{}u'^{}_L ~~~~&&
\nonumber\\
+h_e^{}\bar{e}_R^{}e'^{}_L)-\begin{array}{l}\frac{1}{2}f(\bar{l}_L^c
i\tau_2^{} \xi_L^{} l_L +\bar{l}'^c_R i\tau_2^{} \xi_R^{}
l'^{}_R)\end{array}+\textrm{H.c.}\,~~~&&\nonumber
\end{eqnarray}
\vspace{-1.4cm}
\begin{eqnarray}
\end{eqnarray}
with $h_{d}^{}$, $h_{u}^{}$ and $h_{e}^{}$ being hermitian and $f$
being symmetric. We should keep in mind that the mass terms of the
$SU(2)$-singlet fermions are forbidden by the $Z_2^{}\times Z'^{}_2$
discrete symmetry.

The symmetry breaking pattern should be $SU(2)_L^{}\times
SU(2)_R^{}\times U(1)_{B-L}^{}
\stackrel{\langle\phi_{R}^{}\rangle}{\longrightarrow}
SU(2)_L^{}\times
U(1)_{Y}^{}\stackrel{\langle\phi_{L}^{}\rangle}{\longrightarrow}
U(1)_{em}^{}$, where the vacuum expectation values (VEVs) are
\begin{eqnarray}
\langle\phi_{R(L)}^{}\rangle&=&\langle\phi_{R(L)}^{0}\rangle
=\begin{array}{l}\frac{1}{\sqrt{2}}v_{R(L)}^{}\end{array}\nonumber\\
&&
\begin{array}{l}\textrm{with}~~v_{R(L)}^2\simeq-\frac{\lambda_\phi^{}
\mu_{\phi_{R(L)}^{}}^2-\lambda'^{}_\phi
\mu_{\phi_{L(R)}^{}}^2}{\lambda_\phi^2-\lambda'^2_\phi}\,.\end{array}
\end{eqnarray}
The $SU(2)$-triplet Higgs scalars can pick up the smaller VEVs,
\begin{eqnarray}
&&\begin{array}{l}\langle\xi_{R(L)}^{}\rangle
\end{array}\!\!=\!\!\begin{array}{l}
\langle\xi_{R(L)}^{0}\rangle=u_{R(L)}^{} \simeq
-\frac{\rho_{\xi_{R(L)}^{}}^{}v_{R(L)}^2}{2 M_{\xi_{R(L)}^{}}^2}\ll
v_{R(L)}^{}\end{array}\nonumber\\
&&\begin{array}{l}\textrm{ with}~~
M_{\xi_{R(L)}^{}}^2=\mu_{\xi_{R(L)}^{}}^2+\lambda_{\phi\xi}^{}v_{R(L)}^2
+\lambda'^{}_{\phi\xi}v_{L(R)}^2\,.\end{array}
\end{eqnarray}
In general, the four VEVs are complex and contain four phases.
However, two phases can be eliminated by the $SU(2)_L^{}\times
SU(2)_R^{}\times U(1)_{B-L}^{}$ gauge symmetry while the others can
be set zero by minimizing the scalar potential. So, the four VEVs
are all real. Note we have to require the mass terms
$\mu_{\phi_R^{}}^{2}\neq \mu_{\phi_L^{}}^{2}$, which can arise from
a spontaneous parity violation \cite{cmp1984}, for softly breaking
the parity. Otherwise, we will obtain the unaccepted
$v_R^{}=v_L^{}$. The softly broken parity can also allow
$\mu_{\xi_R^{}}^2\neq \mu_{\xi_L^{}}^2$ and $\rho_{\xi_R^{}}^{}\neq
\rho_{\xi_L^{}}^{}$. For the hierarchical $u_{R(L)}^{}\ll
v_{R(L)}^{}$, most fractions of $\textrm{Im}(\phi_{R,L}^0)$ and
$\phi^{\pm}_{R,L}$ while tiny fractions of
$\textrm{Im}(\xi_{R,L}^0)$ and $\xi^{\pm}_{R,L}$ become the
longitudinal components of the massive gauge bosons $W_{R,L}^\pm$,
$Z'$ and $Z$, which will be clarified later. So, we can simply take
the $SU(2)$-doublet Higgs scalars to be
\begin{eqnarray}
\phi_{R(L)}^{}=\frac{1}{\sqrt{2}}\left[\begin{array}{c}v_{R(L)}^{}+h_{R(L)}^{}\\
[1mm] 0\end{array}\right]\,,
\end{eqnarray}
with the mass terms,
\begin{eqnarray}
V\supset\lambda_\phi^{} v_{R(L)}^2 h_{R(L)}^2+2 \lambda'^{}_\phi
v_L^{} v_R^{} h_L^{}h_R^{}\,.
\end{eqnarray}
As for the real scalar $\chi$, it should have a positive mass term
to respect the unbroken $Z_2^{}\times Z'^{}_2$ discrete symmetry,
\begin{eqnarray}
m_{\chi}^2 =
\mu_{\chi}^2+\lambda_{\chi\phi}[v_{R(L)}^2]+2\lambda_{\chi\xi}[u_{R(L)}^2]>0\,.
\end{eqnarray}

\begin{figure*}
\vspace{5.4cm} \epsfig{file=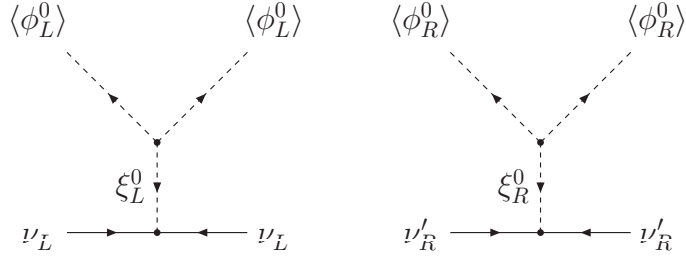, bbllx=6cm, bblly=6.0cm,
bburx=16cm, bbury=16cm, width=8.5cm, height=8.5cm, angle=0, clip=0}
\vspace{-10.3cm} \caption{\label{typeii} The left- and right-handed
type-II seesaw.}
\end{figure*}

After the symmetry breaking, there will be two charged gauge bosons,
\begin{eqnarray}
\begin{array}{l}W^{\pm}_{R(L)}\end{array}&=&\begin{array}{l}\frac{1}{\sqrt{2}}[W^1_{R(L)}\mp i
W^2_{R(L)}]\end{array}~~\textrm{with}\nonumber\\
&&\begin{array}{l}M_{W_{R(L)}^{}}^{2}=\frac{g_2^{2}}{4}[v_{R(L)}^{2}+4u_{R(L)}^2]\end{array}\,,
\end{eqnarray}
two massive neutral gauge bosons,
\begin{eqnarray}
&&\begin{array}{l}Z'=Z_R^{}\cos\alpha-Z_L^{}\sin\alpha\,,~
Z=Z_R^{}\sin\alpha+Z_L^{}\cos\alpha\end{array}\nonumber\\
&&\textrm{with}~~\begin{array}{l}M_{Z',Z}^2=\frac{g_2^{2}[v_{R(L)}^2+8u_{R(L)}^2]
\cos^2_{}\theta_W^{}}{8\cos2\theta_W^{}}\{1\end{array}\nonumber\\
&&\quad\quad\quad\quad\quad\quad~~\begin{array}{l}\left.\pm\sqrt{1-\frac{4(v_L^2+8u_L^2)(v_R^2+8u_R^2)\cos2\theta_W^{}}
{[v_{R(L)}^2+8u_{R(L)}^2]^2_{}\cos^4_{}\theta_W^{}}}\right\}\end{array}\nonumber\\
&&\quad~\,\begin{array}{l}\Rightarrow M_{Z}^{2}\,\simeq
\frac{g_2^2(v_L^2+8u_L^2)}{4\cos^2_{}\theta_W^{}}\,,~
M_{Z'}^{2}\simeq
\frac{g_2^2(v_R^2+8u_R^2)\cos^2_{}\theta_W^{}}{4\cos2\theta_W^{}}
\end{array}\nonumber\\
&&\quad\quad\quad\quad\textrm{for}~~v_L^{2}+8u_L^{2}\ll
v_R^{2}+8u_R^{2}\,,
\end{eqnarray}
and a massless photon,
\begin{eqnarray}
\begin{array}{l}A= B\sqrt{\cos2\theta_W^{}} + W_{L}^3\sin\theta_W^{} +
W_{R}^3\sin\theta_W^{}\,.\end{array}
\end{eqnarray}
Here we have defined \cite{senjanovic1979}
\begin{eqnarray}
\begin{array}{l}\sin^2_{}\theta_W^{}\end{array}&=&\begin{array}{l}\frac{g_1^2}{2g_1^2+g_2^2}\,,\end{array}
\end{eqnarray}
\begin{subequations}
\begin{eqnarray}
\begin{array}{l}Z_R^{}\end{array}&=&\begin{array}{l}-B\tan\theta_W^{}+W_{R}^3\sec\theta_W^{}\sqrt{\cos
2\theta_W^{}}\,,~~~\end{array}\\
\begin{array}{l}Z_L^{}\end{array}&=& \begin{array}{l}-B\tan\theta_W^{}\sqrt{\cos
2\theta_W^{}}+W_{L}^3\cos\theta_W^{}\end{array}\nonumber\\
&&\begin{array}{l}-W_R^3\sin\theta_W^{}\tan\theta_W^{}\,,\end{array}
\end{eqnarray}
\end{subequations}
and
\begin{eqnarray}
\begin{array}{l}\tan2\alpha\end{array}&=&\begin{array}{l}\frac{2\sin^2_{}\theta_W^{}\sqrt{\cos
2\theta_W^{}}}
{\frac{v_R^2+8u_R^2}{v_L^2+8u_L^2}\cos^4_{}\theta_W^{}+\sin^4_{}\theta_W^{}-\cos
2\theta_W^{}}\,.\end{array}
\end{eqnarray}

\begin{figure}
\vspace{3.4cm} \epsfig{file=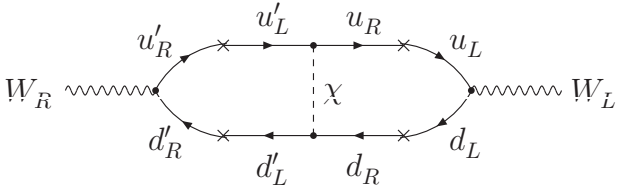, bbllx=4.5cm, bblly=6.0cm,
bburx=14.5cm, bbury=16cm, width=8.5cm, height=8.5cm, angle=0,
clip=0} \vspace{-9.6cm} \caption{\label{wmixing} The two-loop
$W_L^{}-W_R^{}$ mixing.}
\end{figure}

The SM and mirror charged fermions can obtain the Dirac masses
through their Yukawa couplings with the $SU(2)$-doublet Higgs
scalars,
\begin{eqnarray}
\begin{array}{l}m_f^{}=\frac{1}{\sqrt{2}}y_f^{}v_L^{}\,,~~
M_{f'}^{}=\frac{1}{\sqrt{2}}y_f^{}v_R^{}\,,\end{array}
\end{eqnarray}
which yields
\begin{eqnarray}
\label{fmass}
\frac{v_R^{}}{v_L^{}}&=&\frac{M_{d'}^{}}{m_d^{}}=\frac{M_{s'}^{}}{m_s^{}}=\frac{M_{b'}^{}}{m_b^{}}
=\frac{M_{u'}^{}}{m_u^{}}=\frac{M_{c'}^{}}{m_c^{}}=\frac{M_{t'}^{}}{m_t^{}}\nonumber\\
&=&\frac{M_{e'}^{}}{m_e^{}}=\frac{M_{\mu'}^{}}{m_\mu^{}}=\frac{M_{\tau'}^{}}{m_\tau^{}}\,.
\end{eqnarray}
As for the SM left-handed neutrinos and the mirror right-handed
neutrinos, they can obtain the Majorana masses through the left- and
right-handed type-II seesaw \cite{mw1980}, respectively,
\begin{eqnarray}
m_{\nu}^{}=fu_L^{}\,,~~M_{\nu'}^{}=fu_R^{}\,.
\end{eqnarray}
The relevant diagram is shown in Fig. \ref{typeii}. The SM and
mirror neutrino mass eigenvalues should obey
\begin{eqnarray}
\frac{u_R^{}}{u_L^{}}=\frac{M_{\nu'^{}_1}^{}}{m_{\nu^{}_1}^{}}
=\frac{M_{\nu'^{}_2}^{}}{m_{\nu^{}_2}^{}}=\frac{M_{\nu'^{}_3}^{}}{m_{\nu^{}_3}^{}}\,.
\end{eqnarray}
Note that the unitary quark mixing matrix, i.e. the
Cabibbo-Kobayashi-Maskawa \cite{cabibbo1963} (CKM) matrix $V$ and
the unitary lepton mixing matrix, i.e. the Maki-Nakagawa-Sakata
\cite{mns1962} (MNS) matrix $U$ in the SM sector should be
identified with those in the mirror sector. Since the $Z_2^{}\times
Z'^{}_2$ symmetry forbids the mixing between the SM and mirror
fermions, we will not have the one-loop $W_L^{}-W_R^{}$ mixing
\cite{mohapatra1988}, which always exists in the conventional
left-right symmetric models. Instead, the SM and mirror quarks
associated with the real scalar can mediate a two-loop
$W_L^{}-W_R^{}$ mixing as shown in Fig. \ref{wmixing}. We can
estimate the two-loop mass term to be
\begin{eqnarray}
\delta M_{W_L^{}W_R^{}}^2 &\simeq& \frac{3g_2^2 |V_{tb}^{}|^2_{}
h_{u_{33}^{}}^{}h_{d_{33}^{}}^{}
}{2(16\pi^2)^2_{}}\frac{M_{t'}^{}M_{b'}^{}m_t^{}m_b^{}}{\Lambda^2_{}(=M_{t'}^2)}\nonumber\\
&\simeq& \frac{3g_2^2 |V_{tb}^{}|^2_{}
h_{u_{33}^{}}^{}h_{d_{33}^{}}^{}}{2(16\pi^2)^2_{}} m_b^2\ll m_b^2\,,
\end{eqnarray}
which means a negligible $W_L^{}-W_R^{}$ mixing.

The non-perturbative QCD Lagrangian now should be
\begin{eqnarray}
\mathcal{L}_{QCD}^{}\supset\bar{\theta}\frac{g^2_3}{32\pi^2_{}}G\tilde{G}~~\textrm{with}~~
\bar{\theta}=\theta+\textrm{Arg}\textrm{Det} (M_u^{} M_d^{})\,,
\end{eqnarray}
where $\theta$ is the original QCD phase while $M_u^{}$ and $M_d^{}$
are the mass matrices of the SM and mirror down- and up-type quarks,
respectively,
\begin{eqnarray}
\mathcal{L}\supset -[\bar{d}^{}_L,
\bar{d}'^{}_L]M_d^{}\left[\begin{array}{c}d^{}_R\\
d'^{}_R\end{array}\right]-[\bar{u}^{}_L,
\bar{u}'^{}_L]M_u^{}\left[\begin{array}{c}u^{}_R\\
u'^{}_R\end{array}\right]+\textrm{H.c.}&&\nonumber\\
\textrm{with}~ M_d^{}=
\left[\begin{array}{cc}\frac{y_d^{}v_L^{}}{\sqrt{2}}&0\\
0&\frac{y_d^{\dagger}v_R^{}}{\sqrt{2}}\end{array}\right]\,,~M_u^{}=
\left[\begin{array}{cc}\frac{y_u^{}v_L^{}}{\sqrt{2}}&0\\
0&\frac{y_u^{\dagger}v_R^{}}{\sqrt{2}}\end{array}\right]\,.&&
\end{eqnarray}
Clearly, the $\theta$-term should be removed as a result of the
parity invariance. Furthermore, the real determinants
$\textrm{Det}(M_d^{})$ and $\textrm{Det}(M_u^{})$ should induce a
zero $\textrm{Arg}\textrm{Det} (M_u^{} M_d^{})$. We hence can obtain
a vanishing strong CP phase $\bar{\theta}$ \cite{bcs1991}. In the
absence of the mixing between the SM and mirror fermions, the strong
CP phase will keep zero at loop level, unlike that in the universal
seesaw scenario \cite{bm1989}.

\begin{figure*}
\vspace{4.5cm} \epsfig{file=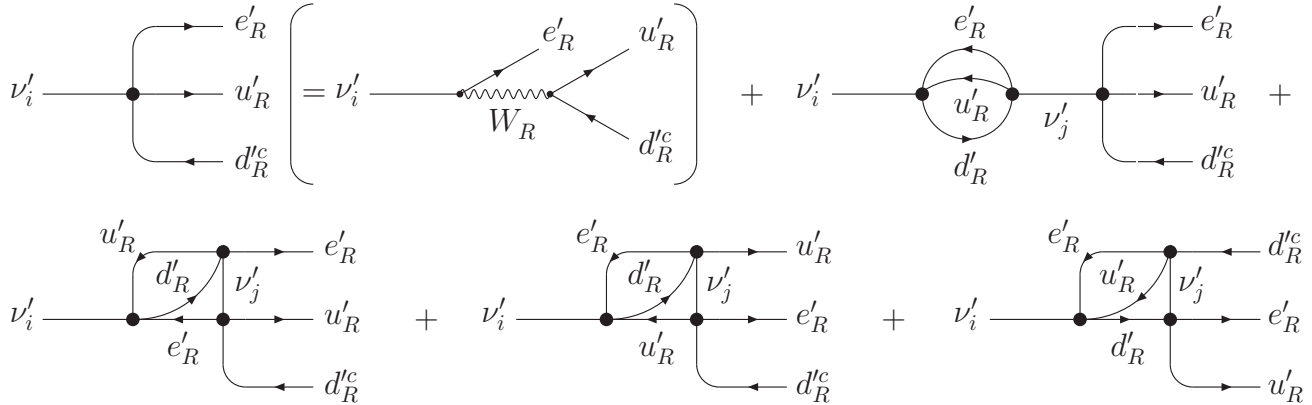, bbllx=5.5cm, bblly=6.0cm,
bburx=15.5cm, bbury=16cm, width=8.5cm, height=8.5cm, angle=0,
clip=0} \vspace{-7.8cm} \caption{\label{mnudecay} The mirror
Majorana neutrinos decay into the mirror charged fermions at
two-loop order. The CP conjugation is not shown for simplicity.}
\end{figure*}

As shown in Fig. \ref{mnudecay}, the right-handed charged gauge
boson can mediate the three-body decays of the mirror neutrinos into
the mirror charged fermions as long as the kinematics is allowed.
For example, if the mirror neutrinos are only heavier than the
first-generation mirror charged fermions, their decay width should
be
\begin{eqnarray}
\Gamma_{\nu'^{}_i}^{}&=&\Gamma_{\nu'^{}_i\rightarrow
e'^{}_{R}+u'^{}_{R}+d'^c_{R}}^{} +\Gamma_{\nu'^{}_i\rightarrow
e'^c_{R}+u'^c_{R}+d'^{}_{R}}^{}\nonumber\\
&\simeq&\frac{g_2^4|V_{ud}^{}|^2_{}|U_{ei}^{}|^2_{}}{2^{10}_{}\pi^3}\frac{M_{\nu'^{}_i}^5}{M_{W_R^{}}^4}
~~\textrm{for}~~M_{\nu'^{}_i}^2\ll M_{W_R^{}}^2\,,
\end{eqnarray}
In the presence of the nonzero CP phases in the unitary MNS matrix,
the mirror neutrino decays can generate a lepton asymmetry in the
mirror charged leptons if the decaying mirror neutrino is not
heavier than the mirror tau lepton. For demonstration, we consider
the case where at least one of the three mirror neutrinos can only
decay into the first-generation mirror charged fermions. The CP
asymmetry can be calculated at two-loop order,
\begin{eqnarray}
\label{cpasymmetry}
\varepsilon_{\nu'^{}_i}^{}&=&\frac{\Gamma_{\nu'^{}_i\rightarrow
e'^{}_{R}+u'^{}_{R}+d'^c_{R}}^{} -\Gamma_{\nu'^{}_i\rightarrow
e'^c_{R}+u'^c_{R}+d'^{}_R}^{}}{\Gamma_{\nu'^{}_i\rightarrow
e'^{}_{R}+u'^{}_{R}+d'^c_{R}}^{} +\Gamma_{\nu'^{}_i\rightarrow
e'^c_{R}+u'^c_{R}+d'^{}_{R}}^{}}\nonumber\\
&=&\left\{\begin{array}{l}\frac{g_2^4
|V_{ud}^{}|^2_{}}{48\pi^3}\sum_{j\neq
i}^{}\frac{\textrm{Im}[(U_{ei}^\ast U_{ej}^{})^2_{}]
}{|U_{ei}^{}|^2_{}}\frac{m_{\nu^{}_i}^{}}{m_{\nu^{}_j}^{}}~~\textrm{for}\\
[1mm] \quad\quad \quad\quad M_{\nu'^{}_{i}}^2\ll
M_{\nu'^{}_{j}}^2\ll
M_{W_R^{}}^2\,,\\
[2mm] \frac{g_2^4 |V_{ud}^{}|^2_{}}{128\pi^3}\sum_{j\neq
i}^{}\frac{\textrm{Im}[(U_{ei}^\ast U_{ej}^{})^2_{}]
}{|U_{ei}^{}|^2_{}}\frac{m_{\nu^{}_i}^{}m_{\nu^{}_j}^{}}{m_{\nu^{}_j}^{2}-m_{\nu^{}_i}^{2}}~~\textrm{for}\\
[1mm] \quad\quad \quad\quad M_{\nu'^{}_{i}}^2\simeq
M_{\nu'^{}_{j}}^2\ll
M_{W_R^{}}^2~~\textrm{and}\\
[1mm] \quad\quad \quad\quad
|M_{\nu'^{}_j}^{2}-M_{\nu'^{}_i}^{2}|^2_{}\gg
M_{\nu'^{}_i}^{}M_{\nu'^{}_j}^{}\Gamma_{\nu'^{}_i}^{}\Gamma_{\nu'^{}_j}^{}\,.
\end{array}\right.
\end{eqnarray}
Since we have known \cite{nakamura2010,stv2011}
\begin{eqnarray}
\begin{array}{l}
\begin{array}{l}
\!\begin{array}{l}V_{ud}^{}=0.97428\pm0.00015\,,~s_{12}^2=0.312^{+0.017}_{-0.015}\,,\end{array}\\
[1mm]\!\begin{array}{l} s_{13}^2=\left\{
\begin{array}{l}0.010^{+0.009}_{-0.006}~\textrm{for~normal~hierarchy}\,,\\
0.013^{+0.009}_{-0.007}~\textrm{for~inverted~hierarchy}\,,\end{array}\right.\end{array}\\
[2mm]
\!\begin{array}{l}\frac{m_{\nu_3^{}}^2-m_{\nu_1^{}}^2}{10^{-3}_{}\,\textrm{eV}^2_{}}=\left\{
\begin{array}{l}2.45\pm 0.09~\textrm{for~normal~hierarchy}\,,\\
-2.34^{+0.10}_{-0.09}~\textrm{for~inverted~hierarchy}\,,\end{array}\right.\end{array}\\
[1mm]
\!\begin{array}{l}\frac{m_{\nu_2^{}}^2-m_{\nu_1^{}}^2}{10^{-5}_{}\,\textrm{eV}^2_{}}=7.59^{+0.20}_{-0.18}\,,~g_2^4=32
G_F^2 M_{W_L^{}}^4=0.182\,,\end{array}
\end{array}\\
\!\begin{array}{l}U_{e1}^{}=c_{12}^{}c_{13}^{}e^{i\frac{\alpha_1^{}}{2}}_{}\,,
\,U_{e2}^{}=s_{12}^{}c_{13}^{}e^{i\frac{\alpha_2^{}}{2}}_{}\,,
\,U_{e3}^{}=s_{13}^{}e^{-i\delta}_{}\,,\end{array}
\end{array}
\end{eqnarray}
the CP asymmetry will only depend on four parameters: one of the
neutrino mass eigenvalues $m_{\nu_{1,2,3}^{}}^{}$, the Dirac CP
phase $\delta$ and the Majorana CP phases $\alpha_{1,2}^{}$. For
example, we can input the best fit values to read
\begin{subequations}
\begin{eqnarray}
\!\!\!\!&&\varepsilon_{\nu'^{}_1}^{}=4.11\times
10^{-7}_{}[\sin(\alpha_2^{}-\alpha_1^{})-0.0057\sin(\alpha_1^{}+2\delta)]\nonumber\\
\!\!\!\!&&~~\textrm{for~the~normal~hierarchy~with}~m_{\nu_1^{}}^{}=10^{-4}_{}\,\textrm{eV}\,,~~\\
\!\!\!\!&&\varepsilon_{\nu'^{}_3}^{}=1.63\times
10^{-7}_{}[\sin(\alpha_1^{}+2\delta)+0.45\sin(\alpha_2^{}+2\delta)]\nonumber\\
\!\!\!\!&&
~~\textrm{for~the~inverted~hierarchy~with}~m_{\nu_3^{}}^{}=10^{-4}_{}\,\textrm{eV}\,.~~
\end{eqnarray}
\end{subequations}

If the left-right symmetry is broken at
$\mathcal{O}(10^{8}_{}\,\textrm{GeV})$, the first-generation mirror
charged fermions can obtain the masses from the weak scale to the
TeV scale. For example, according to \cite{nakamura2010}
\begin{eqnarray}
\label{fmass1} &&m^{}_{e}=0.511\,\textrm{MeV}\,,~m^{}_{u}=
1.7-3.1\,\textrm{MeV}\,,\nonumber\\
&&m^{}_{d}= 4.1-5.7\,\textrm{MeV}\,,~v_L^{}=246\,\textrm{GeV}\,,
\end{eqnarray}
we can expect
\begin{eqnarray}
\label{fmass2} &&M^{}_{e'}=208\,\textrm{GeV}\,,~M^{}_{u'}=
0.69-1.3\,\textrm{TeV}\,,\nonumber\\
&&M^{}_{d'}= 1.7-2.3\,\textrm{TeV}~~\textrm{for}~~v_R^{}=
10^{8}_{}\,\textrm{GeV}\,.
\end{eqnarray}
So, the first-generation mirror charged fermions can be produced at
the ongoing and future colliders due to their couplings with the
gauge bosons and the real scalar. Subsequently, they will decay into
their SM partners with the real scalar being a missing energy,
\begin{eqnarray}
f'\rightarrow f+\chi\,.
\end{eqnarray}
At the same time, we can take
\begin{eqnarray}
&&m^{}_{\nu^{}_{1(3)}}=10^{-4}_{}\,\textrm{eV}\,,~
M^{}_{\nu'^{}_{1(3)}}=10\,\textrm{TeV}~~\textrm{for}\nonumber\\
&&u_L^{}=0.1\,\textrm{eV}\,,~u_R^{}=10^{7}_{}\,\textrm{GeV}\,,
\end{eqnarray}
to generate a desired lepton asymmetry in the mirror electron. As
the mirror electron decays into the SM charged leptons with the real
scalar, the mirror lepton asymmetry can be transferred to the SM
sector. The SM lepton asymmetry eventually can result in a baryon
asymmetry through the sphaleron \cite{krs1985} processes.

A stable SM-singlet scalar can annihilate into the light SM species
through the $s$-channel exchange of the SM Higgs boson. The
annihilation cross section can arrive at about $1\,\textrm{pb}$ to
give a right dark-matter relic density if the dark-matter scalar is
at the weak scale. Meanwhile, the dark-matter scalar can scatter off
the nucleons by the $t$-channel exchange of the SM Higgs boson. The
dark-matter-nucleon scattering can be verified by the forthcoming
dark matter direct detection experimental results. There have been a
lot of works studying such dark-matter scalar \cite{sz1985}. In the
present model, the real scalar $\chi$ definitely is a dark-matter
scalar. However, the mirror charged fermions rather than the SM
Higgs boson can dominate the dark-matter annihilation and
scattering. In particular, when the mirror electron dominates the
dark-matter annihilation, we can detect the dark-matter scalar at
the colliders through the distinguishable decays of the mirror
charged fermions even if the dark-matter-nucleon scattering is not
significant. Furthermore, the mirror charged fermions and the
dark-matter scalar can mediate other interesting processes such as
lepton flavor violation and meson-antimeson mixing. We will study
the mirror charged fermions and the dark-matter scalar in details
elsewhere.

To summarize, we have proposed a mirror left-right symmetric model
with a softly broken parity symmetry. Benefited from the parity
symmetry, the ratio among the mirror charged fermion masses can
equal that among the SM charged fermion masses while the Majorana
mass matrices of the mirror neutrinos and the SM neutrinos can have
a same texture. Our model can solve the strong CP problem without
the axion, like the left-right symmetric model for the universal
seesaw. The mirror charged fermions can decay into their SM partners
with the dark-matter scalar while the mirror neutrinos can decay
into the mirror charged fermions through the right-handed gauge
interactions. We thus can realize a novel leptogensis scenario where
the CP asymmetry is fully determined by the neutrino mass matrix.
Although the left-right symmetry cannot be broken at the accessible
TeV scale, it is possible to examine the mass spectrum of the
first-generation mirror charged fermions at the colliders for a
left-right symmetry breaking scale of the order of
$\mathcal{O}(10^8_{}\,\textrm{GeV})$. Our dark-matter scalar can be
found at the colliders even if it has no significant signal in the
dark matter direct detection experiments.

I thank Manfred Lindner for useful discussions. This work is
supported in part by the Sonderforschungsbereich TR 27 of the
Deutsche Forschungsgemeinschaft.

\end{document}